\documentclass[
  prd,
  twocolumn,
  superscriptaddress,
  nofootinbib,
  amsmath,
  amssymb,
  aps,
  floatfix
]{revtex4-2}

\usepackage{graphicx}
\usepackage{xcolor}
\usepackage{booktabs}
\usepackage{siunitx}
\usepackage{hyperref}
\usepackage{physics}
\usepackage{braket}
\usepackage{subfigure}
\DeclareSIUnit{\mas}{\text{mas}}
\DeclareSIUnit{\arcsec}{\text{arcsec}}
\DeclareSIUnit{\arcmin}{\text{arcmin}}
\DeclareSIUnit{\deg}{\text{deg}}
\DeclareSIUnit{\year}{\text{yr}}
\DeclareSIUnit{\Myr}{\text{Myr}}
\DeclareSIUnit{\Gyr}{\text{Gyr}}
\DeclareSIUnit{\pc}{\text{pc}}
\DeclareSIUnit{\kpc}{\text{kpc}}
\DeclareSIUnit{\Mpc}{\text{Mpc}}

\hypersetup{
  colorlinks=true,
  urlcolor=blue,
  citecolor=green!50!black,
  linkcolor=red!50!black,
}

\usepackage{xcolor}

\begin{document}

\title{Addressing position anomalies in the Strong Gravitational Lensing System HS~0810+2554 through Dark Matter Subhalos}

\author{Yuanlin Gong}
\email{yuanlingong@nnu.edu.cn}
\affiliation{Department of Physics and Institute of Theoretical Physics, Nanjing Normal University, Nanjing, 210023, China}

\author{Lei Wu}
\email{leiwu@njnu.edu.cn}
\affiliation{Department of Physics and Institute of Theoretical Physics, Nanjing Normal University, Nanjing, 210023, China}
\affiliation{Nanjing Key Laboratory of Particle Physics and Astrophysics, Nanjing, 210023, China}

\author{Qiang Yuan}
\email{yuanq@pmo.ac.cn}
\affiliation{Key Laboratory of Dark Matter and Space Astronomy, Purple Mountain
Observatory, Chinese Academy of Sciences, Nanjing 210023, China}
\affiliation{School of Astronomy and Space Science, University of Science and
Technology of China, Hefei 230026, China}

\author{Tengyuan Zhang}
\email{241002029@njnu.edu.cn}
\affiliation{Department of Physics and Institute of Theoretical Physics, Nanjing Normal University, Nanjing, 210023, China}
\affiliation{Key Laboratory of Dark Matter and Space Astronomy, Purple Mountain
Observatory, Chinese Academy of Sciences, Nanjing 210023, China}

\begin{abstract}

Self-bounded dark matter (DM) subhalos are predicted to populate galactic halos in great abundance in the Cold Dark Matter (CDM) scenario. These substructures can leave observable imprints in strong gravitational lensing and have shown the ability to account for flux-ratio and position anomalies in multiply imaged systems. In this paper, we utilize the DM subhalos to address the image position anomalies of the two radio quads of HS 0810+2554 observed with the Very Long Baseline Interferometry. We model the lens using an elliptical power-law macro-lens supplemented by a population of CDM subhalos from numerical simulations and perform a dual-source reconstruction to fit all eight radio images simultaneously. We find that subhalos below $10^{6}M_\odot$ induce astrometric shifts smaller than the measurement uncertainties, whereas more massive subhalos naturally generate the required milliarcsecond perturbations without significantly altering the global lens configuration. Including CDM subhalos improves the fit from $\chi^2=60.38$ for the pure macro-lens to $\chi^2=1.61$. Our results show that the position anomalies of HS~0810+2554 can be explained within the CDM framework and do not by themselves necessarily require non-standard scenarios like fuzzy DM or angular complexity in the macro-lens. Instead, they provide a sharp and testable manifestation of the subhalo population predicted by CDM.

\end{abstract}

\maketitle

\section{Introduction}

Dark Matter (DM) supported by unambiguous observational evidence from cosmic microwave background, large-scale structure surveys and gravitational lensing of galaxies and galaxy clusters is a necessary ingredient in modern cosmology \cite{2005PhR...405..279B,Massey:2010hh,Cirelli:2024ssz}. The standard Cold Dark Matter (CDM) paradigm has achieved remarkable success in the large-scale structure of the Universe \cite{Planck:2018vyg,BOSS:2016wmc}, while the discrepancies on subgalactic scales~\cite{2017ARA&A..55..343B,2017Galax...5...17D,1999ApJ...524L..19M,1999ApJ...522...82K,2011MNRAS.415L..40B} motivate the Fuzzy Dark Matter (FDM) model, an ultralight bosonic particle with a mass of  $10^{-22}$ eV  and a de Broglie wave length of order 1 kpc on galactic scales \cite{Hu:2000ke,Schive:2014dra,Hui:2016ltb,Hui:2021tkt}. These problems could also be the result of baryonic physics at play in galaxy formation and evolution within CDM \cite{Sawala:2009js,Sales:2022ich}. CDM predicts an abundance of subhalos \cite{Gao:2004au,2008MNRAS.391.1685S,2007ApJ...657..262D} on subgalactic scales, populating in the subgalactic scales. The subhalos are greatly suppressed in FDM model owing to the quantum pressure of FDM on distances comparable to its de Broglie wavelength \cite{Schive:2015kza,May:2021wwp}. Therefore, understanding the nature and distribution of subgalactic DM structures \cite{Irsic:2017yje,Armengaud:2017nkf,Hsueh:2019ynk,Vegetti:2023mgp,Gilman:2026uvq} provides an important avenue for testing the validity of DM models and probing the underlying physics of DM \cite{2018Natur.562...51B}. Supermassive Black Hole evolution in the  high-redshift \cite{Ellis:2025dpw}, star formation rate \cite{Urrutia:2025fvp}, and gravitational lensing of the galaxies and quasar \cite{Mao:1997ek,Chiba:2001wk,Dalal:2001fq,Amara:2004dr,Koopmans:2005nr,Li:2016afu,Lei:2025pky} have provided such discriminators of DM scenarios by detecting subhalos.


Subhalos hosted by lensing galaxies and the line-of-sight halos can manifest themselves in strong gravitational lensing by perturbing macro-lens potentials and altering image magnifications, time delays, image positions and extended arcs \cite{1992grle.book.....S,2010ARA&A..48...87T,2010CQGra..27w3001B}. Many works have used them to explain the flux ratio \cite{Mao:1997ek,Metcalf:2001ap,Dalal:2001fq,Mao:2004iw,Xu:2009ch,Xu:2014dda}, time delay~\cite{Keeton:2008gq,Congdon:2009fa,Liao:2018ofi,Vujeva:2025nwg} and image position \cite{Metcalf:2001ap,Chen:2006ym,Chen:2008vt,Xu:2009ch,Nierenberg:2014cga,Inoue:2012px,Hsueh:2019ynk,Gilman:2022ida,Abe:2023hlf} anomalies in the quadruply-lensed quasars or the galaxy-galaxy  strong gravitational lensing systems. 
While some works refer to the baryonic structures, such as stellar discs \cite{Hsueh:2016aih,Hsueh:2017nlk,Hsueh:2017zfs,Gilman:2016uit}, and complexity \cite{2006MNRAS.370.1339H,2018MNRAS.475.1987G,2022MNRAS.516.1808P,2024A&A...688A.110S} in the macro-lens to explain them. Moreover, due to the wave nature of FDM model, the mass density profiles of DM haloes exhibit small-scale fluctuations termed as “granules” due to wave interference \cite{Schive:2014dra,Schive:2015kza,May:2021wwp}, which is vastly different from the haloes expected in the CDM. It has therefore been proposed that FDM-induced density fluctuations may resolve flux-ratio and position anomalies in certain quadruply lensed systems \cite{Chan:2020exg,Hou:2026tsl}.

Previous studies have shown that FDM-induced wave perturbations can successfully explain the position anomalies observed in the two radio quads of HS~0810+2554 \cite{2015MNRAS.454..287J,2014ApJ...783...57C,2016ApJ...824...53C,2019MNRAS.485.3009H}, provided that the FDM particle mass is of order $10^{-22}$ eV \cite{2023NatAs...7..736A}. But such mass is opposed by Jeans modeling of dwarf spheroidal galaxies \cite{Chen:2016unw,Safarzadeh:2019sre}, the power spectrum of the Lyman-$\alpha$ Forest \cite{Irsic:2017yje,Armengaud:2017nkf}, the stellar velocity dispersions in ultra-faint dwarf galaxies \cite{Dalal:2022rmp}, supermassive Black Hole evolution in the  high-redshift \cite{Ellis:2025dpw} and so on \cite{Banik:2019smi,DES:2020fxi,Rogers:2020ltq,Powell:2023jns}. More recently, the multipoles in the form of lopsidedness, triangleness, boxiness and disciness have also been shown to provide an adequate explanation \cite{2025MNRAS.543.3952M}. These results demonstrate that HS~0810+2554 is highly sensitive to small-scale or angular complexity in the lens potential. However, an important question remains unanswered: can the position anomalies observed in HS~0810+2554 be naturally explained by the subhalo populations already predicted by the standard CDM paradigm? Addressing this question is essential for evaluating whether non-standard DM models are genuinely required by current observations or whether the observed discrepancies can be understood within the conventional cosmological framework.

In this work, we develop a hierarchical lens-modeling framework for HS~0810+2554, progressing from a smooth macro-lens to increasingly realistic descriptions of DM substructure. We first construct a smooth Elliptical Power-Law (EPL) lens model and subsequently investigate the influence of both an individual Navarro--Frenk--White (NFW) subhalo and a cosmologically motivated subhalo population generated according to the Aquarius simulations \cite{2008MNRAS.391.1685S}. The subhalo population is generated following the predicted subhalo mass function and spatial distribution, with a physically motivated lower mass limit determined from their expected astrometric impact. This hierarchical modeling framework provides a systematic approach for assessing whether the observed position anomalies in HS~0810+2554 can be explained within the standard CDM paradigm.

\section{LENS MODELING WITH DM SUBHALOS}
\noindent
In this section, we present the complete methodological framework adopted to model the strong gravitational lens system HS 0810+2554 and to investigate the role of DM substructure in erasing the observed position anomalies. We begin with an overview of the observational characteristics of the system, emphasizing its unique octuple-image configuration revealed by high-resolution European Very Long Baseline Interferometry (VLBI) Network radio observations. We then introduce the theoretical framework of smooth lens modeling, in which the foreground galaxy is described by an Elliptical Power-Law (EPL) mass profile supplemented by external shear—the combination referred to as the EPL plus external shear model (or, hereafter, the smooth lens model)—and the dual-source structure of the background quasar is incorporated to reproduce the observed image multiplicity. Finally, we extend this baseline model by including DM subhalos following the Aquarius simulations motivated subhalo mass function and spatial distribution, enabling a systematic assessment of their impact on image positions. 

\subsection{Observational Overview of HS 0810+2554}
\label{subsec:observational_overview}

HS~0810+2554 is a well-studied galaxy-scale strong gravitational lens system originally identified in optical observations with the {Hubble Space Telescope} (HST), where it appears as a quadruply imaged quasar lens~\cite{2002A&A...382L..26R}. The system is often described as a ``bright twin'' of PG~1115+080 due to its similar fold-image configuration. The foreground lens galaxy lies at a redshift of $z_l \simeq 0.89$, while the background source is a radio-quiet quasar located at $z_s = 1.51$~\cite{2014ApJ...783...57C}. Subsequent high-resolution radio observations revealed that HS~0810+2554 possesses a significantly more complex structure than suggested by its optical appearance. Deep e-MERLIN and European VLBI Network observations uncovered two compact nano-Jansky radio-emitting components associated with the background quasar, likely corresponding to a radio core and a jet-like feature~\cite{2015MNRAS.454..287J,2019MNRAS.485.3009H}. Each radio component is independently lensed by the foreground galaxy, producing two additional quadruply imaged systems. As a result, HS~0810+2554 contains three distinct quads originating from the same background source: one optical quad and two radio quads.

The discovery of the radio quads motivated a series of increasingly sophisticated lens-modeling studies. Early analyses based on the optical images demonstrated that smooth elliptical mass distributions combined with external shear could broadly reproduce the overall lens geometry~\cite{2020MNRAS.496..598C,Nierenberg:2019pdj}. Following the detection of the radio structure, Hartley et al.~\cite{2019MNRAS.485.3009H} modeled the eight VLBI image positions using conventional smooth lens models and found that significant astrometric residuals remained. Unlike optical observations, radio interferometric measurements in HS~0810+2554 are primarily sensitive to compact jet emission rather than extended quasar light. However, the lack of short baselines in VLBI observations can lead to incomplete recovery of extended flux, introducing systematic uncertainties in the measured flux ratios~\cite{2019MNRAS.485.3009H}. In contrast, the image positions are measured with milliarcsecond (mas) precision and are considerably less affected by such observational limitations. For this reason, recent lens-modeling studies have generally relied on the radio astrometry while excluding flux-ratio information from the fitting procedure~\cite{2019MNRAS.485.3009H,2023NatAs...7..736A,2025MNRAS.543.3952M}.

The inferred discrepancies could not be fully eliminated through adjustments of the smooth mass distribution alone, suggesting the presence of additional small-scale perturbations in the lens potential. Recently, Amruth et al.~\cite{2023NatAs...7..736A} further confirmed that various standard smooth models, including Singular Isothermal Ellipsoid and EPL profiles with external shear, are unable to simultaneously reproduce all observed image positions. Therefore, based on the granules structure, they demonstrated that the FDM can elegantly resolve the position anomalies with appropriate mass of FDM particle~\cite{2023NatAs...7..736A}.
Also, to address this tension, Miller et al.~\cite{2025MNRAS.543.3952M} explored more flexible mass models incorporating higher-order angular perturbations and showed that such extensions can substantially improve the fit to the radio astrometry. Despite these advances, the physical origin of the observed position anomalies remains uncertain. Existing studies have primarily focused on modifications of the smooth lens potential, including higher-order angular perturbations and resort to other DM paradigm. It leaves an explicitly modeling a cosmologically motivated DM subhalo population remains untested. Consequently, whether the observed astrometric anomalies can be explained within the standard CDM framework remains an open question.


In this work, we investigate whether the position anomalies observed in HS~0810+2554 can be explained by a DM subhalo population predicted by the Aquarius simulations~\cite{2008MNRAS.391.1685S}. Rather than introducing phenomenological perturbations to the lens potential, we construct physically motivated realizations of CDM substructure based on cosmological simulations and quantify their impact on the observed VLBI image positions. This approach enables a direct test of whether the small-scale subhalos expected in the CDM paradigm are sufficient to account for the observed astrometric anomalies.

\subsection{Theoretical Framework for  Lens Modeling}\label{sec:IIB}

The strong gravitational lens system HS~0810+2554 exhibits eight radio images with mas astrometric precision, providing stringent constraints on the mass distribution of the foreground lens galaxy. To describe the smooth mass distribution of the lens galaxy, we adopt an EPL profile together with an external shear component. The EPL profile describes the smooth mass distribution of the lens galaxy, while the external shear accounts for perturbations induced by the surrounding environment. External shear effectively describes the cumulative gravitational influence of nearby galaxies, galaxy groups, and large-scale structures that are not explicitly included in the lens model~\cite{2014ApJ...783...57C,2016ApJ...824...53C}. Such perturbations can modify image positions, magnifications, flux ratios, and the shapes of critical curves and caustics. Previous studies have shown that EPL plus external shear models fail to reproduce all observed image positions in HS~0810+2554~\cite{2019MNRAS.485.3009H}. Nevertheless, we will first establish a baseline model for subsequent analysis by optimizing the smooth lens model, before including the subhalos in the lensing potential. 

The projected convergence of the EPL profile is given by~\cite{Tessore:2015baa}
\begin{equation}
\kappa_{\rm EPL}(R)=\frac{3-\gamma}{2}\left(\frac{\theta_E}{R}\right)^{\gamma-1},
\label{eq:epl_kappa}
\end{equation}
where $\theta_E$ is the Einstein radius, $\gamma$ is the logarithmic three-dimensional density slope, and $R$ denotes the elliptical radius measured from the lens center. The elliptical radius is defined directly in the image plane coordinates as $R=\sqrt{q^2(x-x_{\rm lens})^2+{(y-y_{\rm lens})^2}}$ where $q$ is the axis ratio of the projected mass distribution, defined as the ratio of the minor axis to the major axis, and $(x_{\rm lens},y_{\rm lens})$ denotes the lens center. The lens center $(x_{\rm lens},y_{\rm lens})$ is identified with the centroid of the lens galaxy in the smooth model. In this work, the lens modeling is performed using \texttt{lenstronomy} package~\cite{2018PDU....22..189B} and the lens ellipticity is parameterized by the ellipticity amplitude, which is related to $q$ by $e=\sqrt{e_1^2 + e_2^2}={(1-q)}/{(1+q)}$. Therefore, the EPL model is fully specified by the Einstein radius $\theta_E$, the density slope $\gamma$, the lens center $(x_{\rm lens},y_{\rm lens})$, and the ellipticity components $(e_1,e_2)$. The lensing potential associated with the EPL profile satisfies the two-dimensional Poisson equation
\begin{equation}
\nabla^2 \psi_{\rm EPL}=2\kappa_{\rm EPL}.
\label{eq:epl_poisson}
\end{equation}
However, for general elliptical power-law mass distributions, no simple closed-form expression exists for the corresponding lensing potential. Therefore, instead of explicitly solving Eq.~(\ref{eq:epl_poisson}), we can compute the deflection field using the semi-analytic formalism developed in~\cite{Tessore:2015baa}, which directly evaluates the EPL deflection angles from the projected surface mass distribution. This implementation is adopted in \texttt{lenstronomy}, ensuring both computational efficiency and accuracy.

The lensing potential of the external shear is given by
\begin{equation}
\psi_{\rm ES}(r,\theta)=\frac{\gamma_{\rm ext}}{2}r^2\cos\left[2(\theta-\theta_{\rm ext})\right],
\label{eq:shear_potential}
\end{equation}
where $\gamma_{\rm ext}$ denotes the shear strength and $\theta_{\rm ext}$ specifies its orientation. In the \texttt{lenstronomy} implementation, the external shear is parameterized by the Cartesian components $(\gamma_1,\gamma_2)$, which are related to $(\gamma_{\rm ext},\theta_{\rm ext})$ through $\gamma_1=\gamma_{\rm ext}\cos(2\theta_{\rm ext}), \,
\gamma_2=\gamma_{\rm ext}\sin(2\theta_{\rm ext})$.
Owing to the linearity of the lensing potential, the total smooth potential can be written as
\begin{equation}
\psi_{\rm smooth}=\psi_{\rm EPL}+\psi_{\rm ES}.
\label{eq:smooth_potential}
\end{equation}
The corresponding total deflection angle is given by
\begin{equation}
\boldsymbol{\alpha}_{\rm smooth}=\nabla \psi_{\rm smooth}.
\end{equation}
Finally, the image configuration is determined through the lens equation
\begin{equation}
\boldsymbol{\beta}=\boldsymbol{\theta}-\boldsymbol{\alpha}_{\rm smooth}(\boldsymbol{\theta}),
\label{eq:lens_equation}
\end{equation}
where $\boldsymbol{\beta}$ and $\boldsymbol{\theta}$ denote the angular positions of the source and image in the source plane and image plane, respectively. Solving Eq.~(\ref{eq:lens_equation}) yields the predicted image positions, which can be directly compared with the observed VLBI astrometric data to constrain the lens model parameters.


Aside from the aforementioned free parameters of the lens model, the source configuration is described by the source coordinates $(\beta_{x1},\beta_{y1})$ and the relative angular separation parameter $\theta_{\rm rel}$ inferred from the observed radio morphology. The model parameters are constrained by minimizing the positional offsets between the observed and predicted image coordinates. The corresponding average positional chi-squared statistic is defined as
\begin{multline}
\chi^{2}
=
\frac{1}{16}
\sum_{j=1}^{2}
\sum_{i=1}^{4}
\Biggl[
\frac{
\left(
\theta_{x,ij}^{\rm obs}
-
\theta_{x,ij}^{\rm mod}
\right)^2
}
{\sigma_{x,ij}^2}
+
\frac{
\left(
\theta_{y,ij}^{\rm obs}
-
\theta_{y,ij}^{\rm mod}
\right)^2
}
{\sigma_{y,ij}^2}
\Biggr],
\label{eq:chi2_pos}
\end{multline}
where $\theta_{x,ij}^{\rm obs}$ and $\theta_{y,ij}^{\rm obs}$ denote the observed VLBI image coordinates taken from Table 5 of~\cite{2019MNRAS.485.3009H}, while $\theta_{x,ij}^{\rm mod}$ and $\theta_{y,ij}^{\rm mod}$ are the corresponding model predictions. The quantities $\sigma_{x,ij}$ and $\sigma_{y,ij}$ represent the astrometric uncertainties~\cite{2019MNRAS.485.3009H}, ranging from $\pm$0.8 to $\pm$4.1 mas. The normalization factor $1/16$ ensures that the statistic corresponds to the average positional mismatch per image. The posterior probability distribution is sampled using the affine-invariant MCMC algorithm implemented in the {\tt emcee} package.

\begin{figure}[t]
\centering
\includegraphics[width=0.445\textwidth]{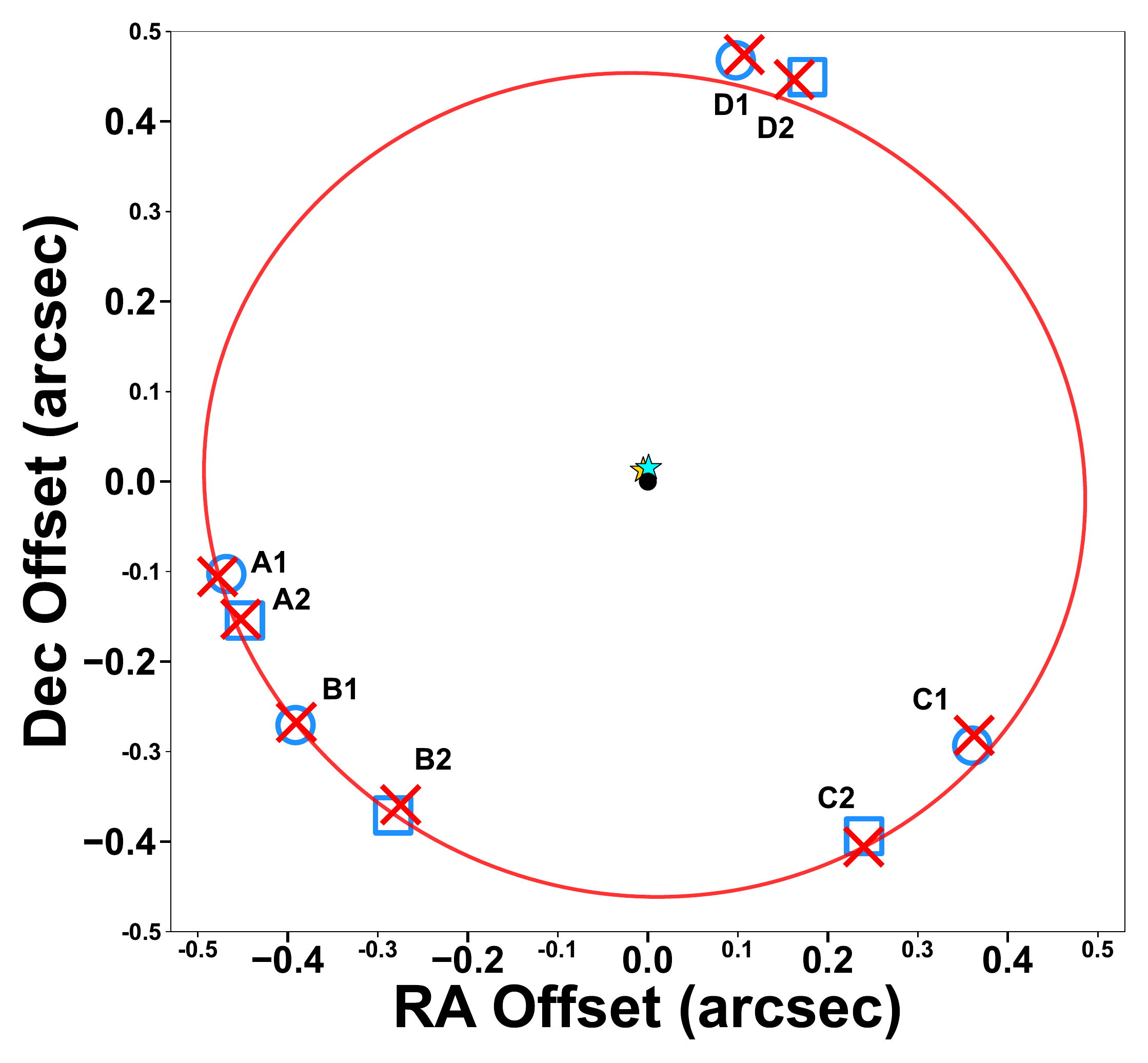}
\caption{Lens geometry of HS 0810+2554 from the smooth dual-source model. 
The coordinate system is centered on the lens galaxy, with the lens center defined as the origin, marked by a black point. The reconstructed configuration shows the image plane with the tangential critical curve (red circle). The eight observed VLBI image positions (red crosses) are compared with the best-fit predicted image positions of the smooth lens model (blue squares for Source 1, blue circles for Source 2). Residual offsets between the observed and model predictions yield a $\chi^2$ value of 60.38, quantifying the position anomalies unresolved by the smooth mass distribution.}
\label{fig:lens_config}
\end{figure}

The prior ranges (the second column) and best-fit values (the third column) for the free parameters of the smooth model obtained from the optimization are summarized in Table~\ref{tab:mcmc_parameters}. These parameters define the best-fit smooth lens model and provide the reference solution against which the impact of DM subhalos is subsequently evaluated. The reconstructed image configuration of the best-fit smooth model is shown in Figure~\ref{fig:lens_config}. The image plane geometry includes a red circular contour representing the critical curve, where strong lensing occurs. The eight observed image positions measured through VLBI observations are marked by red crosses, while the best-fit predicted image positions of the smooth lens model are represented by blue squares (Source 1) and blue circles (Source 2). The center of the lens galaxy is indicated by a black star. The model successfully reproduces the overall geometry of the lens system, including the arrangement of the eight lensed images around the critical curve. The best-fit Einstein radius is $\theta_E=0.4742''$ and the  corresponding host halo mass is $M_{\mathrm{host}}= 1.785\times10^{12}\,M_\odot$,  consistent with the values reported in~\cite{2019MNRAS.485.3009H,2023NatAs...7..736A,2025MNRAS.543.3952M}. However, noticeable offsets remain between the observed VLBI image positions and the model predictions. These residual astrometric discrepancies result in a positional statistic of $\chi^2 = 60.38$, indicating that the EPL plus external shear model cannot fully account for the observed image positions. The remaining positional offsets suggest the presence of additional small-scale perturbations that are not captured by a smooth mass distribution alone. Motivated by this discrepancy, we next introduce a DM subhalo population generated according to the Aquarius simulations and investigate their impact on the image positions of HS~0810+2554.

\begin{table}[htbp]
\centering
\caption{Best-fit values of the 11 free parameters (the first column) obtained from the MCMC optimization for the smooth-lens model (the third column) and macro-lens of the single-subhalo model (the fourth column). In both models, the foreground lens galaxy is modeled by an EPL profile plus external shear; the EPL profile is characterized by $\theta_E$, the logarithmic density slope $\gamma$, the lens-center coordinates $(x_{\rm lens},y_{\rm lens})$, and the ellipticity components $(e_1,e_2)$. The external lens environment is described by the shear components $(\gamma_1,\gamma_2)$. The source configuration is parameterized by the position of the first radio component $(\beta_{x1},\beta_{y1})$ and the relative angular separation $\theta_{\rm rel}$ between the two source components. The second column shows the prior range for each free parameter. See more details in main text.}
\label{tab:mcmc_parameters}
\begin{tabular}{cccc}
\toprule
Parameter & Prior Range & Smooth & Single-subhalo \\
\midrule
$\theta_E$ (arcsec) 
& {\small$\mathcal{U}(0,1)$}
& 0.4742 
& 0.4610 \\[4pt]

$\gamma$
& {\small$\mathcal{U}(1.5,2.5)$}
& 2.3009 
& 2.2940 \\[4pt]

$x_{\rm lens}$ (arcsec)
& {\small$\mathcal{U}(-1,1)$}
& 0.0780 
& 0.0787 \\[4pt]

$y_{\rm lens}$ (arcsec)
& {\small$\mathcal{U}(-1,1)$}
& 0.1056 
& 0.1050 \\[4pt]

$e_1$
& {\small$\mathcal{U}(-0.5,0.5)$}
& 0.0832 
& 0.0842 \\[4pt]

$e_2$
& {\small$\mathcal{U}(-0.5,0.5)$}
& -0.0161 
& -0.0155 \\[4pt]

$\gamma_1$
& {\small$\mathcal{U}(-0.1,0.1)$}
& 0.0276 
& 0.0268 \\[4pt]

$\gamma_2$
& {\small$\mathcal{U}(-0.1,0.1)$}
& 0.0074 
& 0.0072 \\[4pt]

$\beta_{x1}$ (arcsec)
& {\small$\mathcal{U}(-1,1)$}
& 0.0728 
& 0.0744 \\[4pt]

$\beta_{y1}$ (arcsec)
& {\small$\mathcal{U}(-1,1)$}
& 0.1181 
& 0.1202 \\[4pt]

$\theta_{\rm rel}$ (rad)
& {\small$\mathcal{U}(0,2\pi)$}
& 0.4367 
& 0.4389 \\
\bottomrule
\end{tabular}
\end{table}

\subsection{Implementation of the Subhalo Population}\label{sec:IIec}

In this subsection, we describe how to construct a DM subhalo population based on the Aquarius simulations~\cite{2008MNRAS.391.1685S}. The Aquarius simulations provide high-resolution cosmological realizations of six Milky Way-sized DM halos with five levels of numerical resolution and offer detailed predictions for both the abundance and spatial distribution of subhalos within host halos. 
The numbers of subhalos with mass $M$ are generated according to the Aquarius subhalo mass function~\cite{2008MNRAS.391.1685S},
\begin{equation}
\frac{dN}{dM} = a_0 \left(\frac{M}{m_0}\right)^{-1.9},
\label{eq:mass_function}
\end{equation}
where $m_0$ denotes the reference subhalo mass scale of the mass function, which serves to non-dimensionalize the subhalo mass $M$, and $a_0$ denotes the normalization constant controlling the overall abundance of subhalos. In the Aquarius simulations, the subhalo mass function was calibrated for Milky Way-sized host halos with virial masses of approximately $(0.8\text{--}1.8)\times10^{12}\,M_\odot$. For the Aq-D-2 halo in the simulations with Level-2 numerical resolution, Springel et al.~\cite{2008MNRAS.391.1685S} obtained a normalization of $a_{0,\rm Aq}=3.259\times10^{-5}\,M_\odot^{-1}$ and $m_0=2.519\times10^{7}\,M_\odot$ corresponding to a host halo mass of $M_{\rm Aq}=1.774\times10^{12}\,M_\odot$. Given the proximity between the best-fit host halo mass $M_{\mathrm{host}}= 1.785\times10^{12}\,M_\odot$ in the smooth lens model and the $M_{\rm Aq}$, we use $a_{0,\rm Aq}$ in our work.

We allow the host halo mass $M_{\rm host}$ to vary in our analysis and the normalization of the subhalo mass function is accordingly rescaled by,
\begin{equation}
a_0 =a_{0,\rm Aq} \left( \frac{M_{\rm host}}{M_{\rm Aq}} \right).
\label{eq:a0_scaling}
\end{equation}
This scaling preserves the Aquarius prediction for the relative mass fraction of subhalos while allowing the subhalo population to remain consistent with variations in the host halo mass explored in this work. In this way, we do not introduce an additional free parameter to optimize the lens modeling and the subhalo population depends only on host halo mass, or equivalently on the Einstein radius of the EPL model.

\begin{figure}[t]
\centering
\includegraphics[width=1.05\columnwidth]{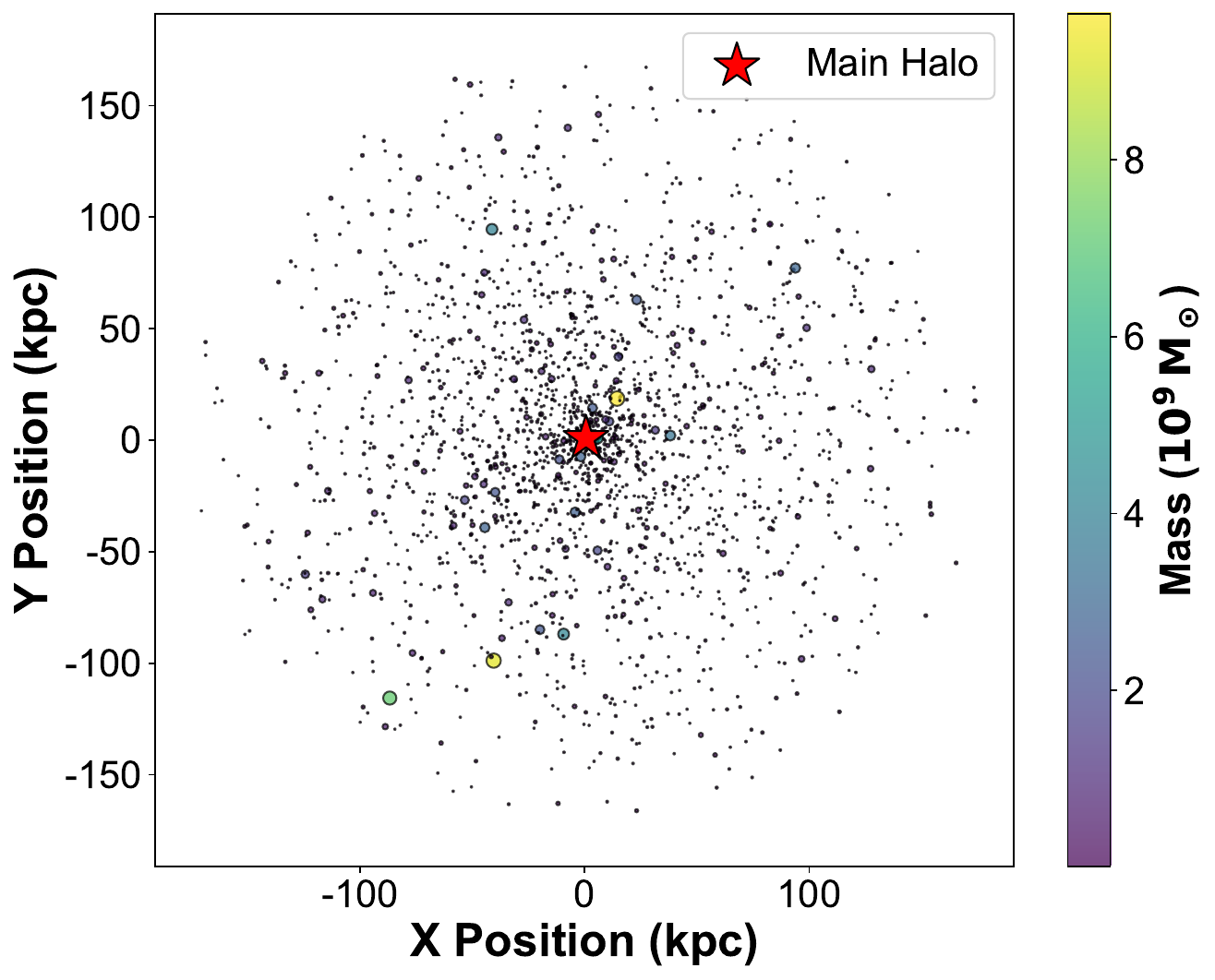}
\caption{
Projected spatial distribution of CDM subhalos in the lens plane of HS~0810+2554.
The host halo mass is $M_{\rm host}=1.43\times10^{12}M_\odot$. The color bar indicates the masses of subhalos with $M_{\rm sub}>10^{9}M_\odot$, while lower-mass subhalos are represented by black points. The red star denotes the center of the host halo.
}
\label{fig:subhalo_distribution}
\end{figure}

To investigate the contribution of subhalos with different masses, the full mass range from $10^{10}~M_\odot$ to $10^{5}\,M_\odot$ considered in this work is divided into logarithmic mass intervals, each spanning one decade in mass as commonly adopted. Treating each interval separately enables us to generate the corresponding subhalo population according to the mass function and to quantify the astrometric perturbations produced by different mass scales. For each subhalo mass interval considered, the expected number of subhalos is first obtained by integrating Eq.~(\ref{eq:mass_function}) over the corresponding mass range. The expectation value is then converted into an integer and adopted as the total number of subhalos generated within that mass interval. Subsequently, the subhalo mass function described by Eq.~(\ref{eq:mass_function}) is normalized and used as the probability density function for the subhalo masses, from which individual subhalo masses $M_i$ are randomly sampled. 
The spatial distribution of subhalos with given mass interval is modeled as the Einasto profile \cite{2008MNRAS.391.1685S},
\begin{equation}
\rho(r) = \rho_{-2}
\exp
\left[
-\frac{2}{\alpha}
\left(
\left(\frac{r}{r_{-2}}\right)^\alpha
-1
\right)
\right],
\label{eq:einasto}
\end{equation}
where $\rho_{-2}$ is the density at radius $r_{-2}$, defined as the location where the logarithmic slope of the profile equals $-2$, and $\alpha$ controls the shape of the profile. Following the Aquarius simulations, we adopt $r_{-2}=199~{\rm kpc}$ and $\alpha=0.678$~\cite{2008MNRAS.391.1685S}. Radial positions $r_i$ are sampled from the corresponding cumulative distribution function, while angular coordinates are drawn from isotropic distributions. The three-dimensional positions are projected onto the lens plane, and the projected coordinates define the centers $(x_i,y_i)$ of the generated subhalos. The resulting projected subhalo distribution for selected host halo mass $M_{\rm host}=1.43\times10^{12}M_\odot$ in the lens plane is illustrated in Figure~\ref{fig:subhalo_distribution}. Massive subhalos are color coded according to their masses, whereas lower-mass subhalos are shown as black points. The red star marks the center of the host halo.

Each generated subhalo is modeled by a NFW density profile~\cite{1996ApJ...462..563N},
\begin{equation}
\rho(r)=
\frac{\rho_{s,i}}
{\left(r/r_{s,i}\right)\left(1+r/r_{s,i}\right)^2},
\end{equation}
where $r_{s,i}$ and $\rho_{s,i}$ denote the scale radius and characteristic density that are directly related to the mass $M_i$ of the $i$th subhalo, respectively. 
Projecting the NFW density profile onto the lens plane yields the convergence of the $i$th subhalo,
\begin{equation}
\kappa_i^{\rm NFW}(R_i)
=
2\frac{\rho_{s,i}r_{s,i}}
{\Sigma_{\rm crit}}
F\left(\frac{R_i}{r_{s,i}}\right),
\end{equation}
where $R_i=\sqrt{(x-x_i)^2+(y-y_i)^2}$ is the projected distance from the center of the subhalo, $\Sigma_{\rm crit}$ is the critical surface mass density, and $F(x)$ is given by Golse and Kneib~\cite{2002A&A...390..821G}.

The convergence thus the lensing potential contributed by the entire subhalo population is obtained by summing over all generated subhalos and added to the total smooth potential $\psi_{\rm smooth}$ as described in Section~\ref{sec:IIB}. The corresponding deflection angles and image positions are subsequently computed using the numerical routines implemented in \texttt{lenstronomy}, which automatically evaluates the lensing quantities of the combined mass distribution and solves the lens equation for the updated lens model.

\begin{figure}[t]
\centering
\includegraphics[width=0.5\textwidth]{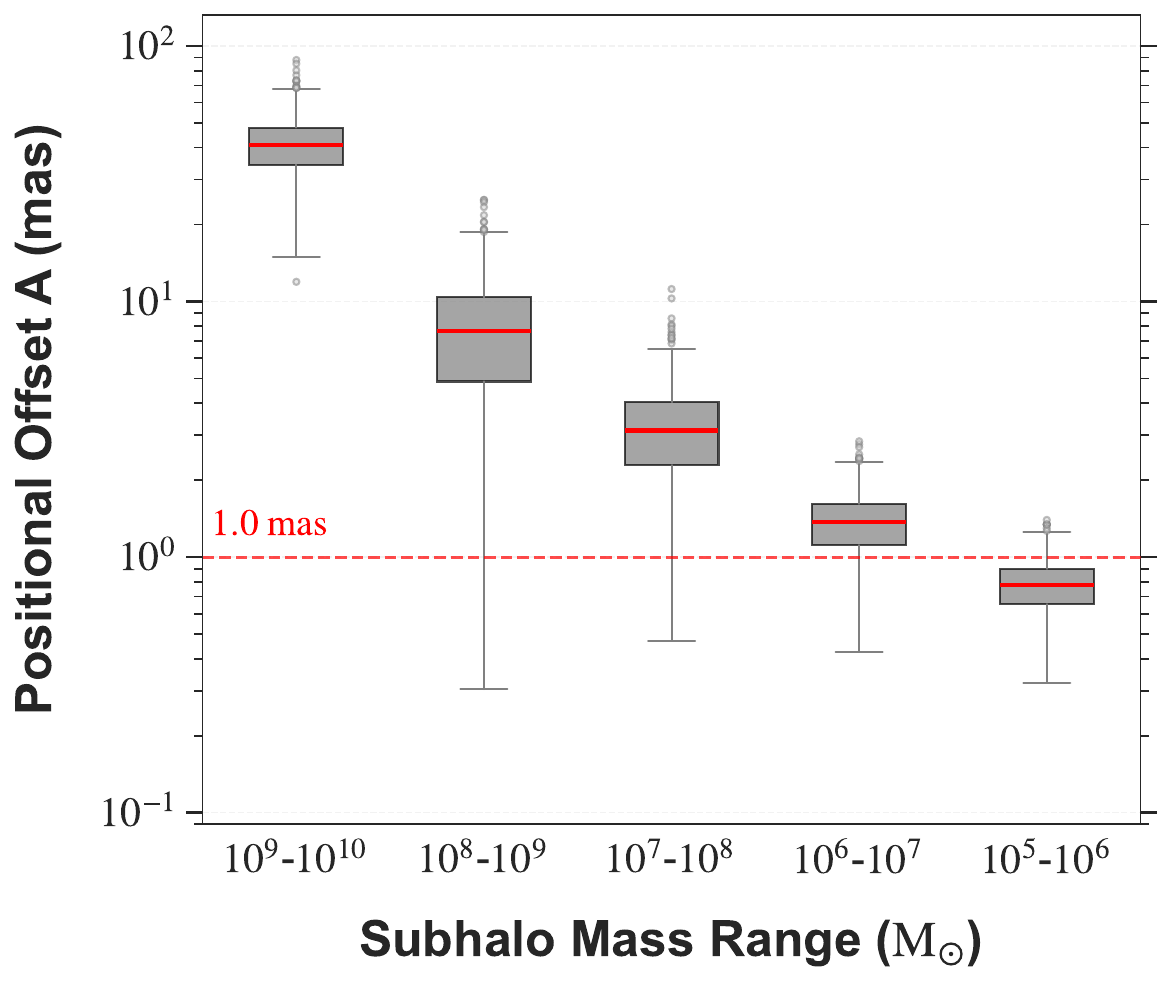}
\caption{The distribution of the positional offset $A$ induced by subhalo populations across five subhalo mass intervals (from $10^{9}$--$10^{10}\,M_{\odot}$ to $10^{5}$--$10^{6}\,M_{\odot}$), based on 1000 realizations per mass interval. The vertical gray lines (whiskers) extend to the furthest data points within 1.5 times the interquartile range (gray band), while individual outliers beyond this range are shown as open circles, indicating rare realizations with exceptionally large positional perturbations. The central red solid line indicates the median offset, and the bottom and top edges of the gray band represent the 25\% and 75\% quartiles. Both the median offset and the interquartile range decrease systematically with decreasing subhalo mass.
The horizontal red dashed line marks the typical VLBI astrometric uncertainty of $\sim\!1\,\mathrm{mas}$, which serves as an observational detectability threshold. 
}
\label{fig:mass_truncation_criterion}
\end{figure}

Although the Aquarius model predicts subhalos over a broad mass range, the gravitational influence of low-mass subhalos may fall below the astrometric sensitivity of current strong-lensing observations. Including such perturbers would substantially increase the computational cost of the simulations while contributing negligibly to the predicted image positions. We therefore establish a physically motivated lower mass threshold before constructing the final subhalo realizations. To quantify the lensing impact of subhalos with different masses, we perform a systematic investigation using Aquarius-based subhalo populations generated within successive mass intervals spanning $10^{9}$--$10^{10}\,M_\odot$ down to $10^{5}$--$10^{6}\,M_\odot$. As mentioned above, in each interval, subhalo masses are sampled according to the Aquarius mass function, while their spatial distribution follows the Einasto profile. The resulting subhalo populations therefore preserve the statistical properties predicted by the Aquarius simulations while isolating the contribution from different mass scales. To characterize the perturbation strength, we define the positional offset $A$ as the root-mean-square displacement between image positions predicted by lens models with and without subhalos,
\begin{equation}
A^{2} = \frac{1}{4}
\sum_{i=1}^{4}
\left[
(x_{\mathrm{sub},i}-x_{\rho,i})^{2}
+
(y_{\mathrm{sub},i}-y_{\rho,i})^{2}
\right],
\end{equation}
where $(x_{\mathrm{sub},i},y_{\mathrm{sub},i})$ denotes the position of the $i$-th lensed image predicted by the model with subhalo perturbations, and $(x_{\rho,i},y_{\rho,i})$ denotes the corresponding image position predicted by the smooth lens model. The summation extends over the four lensed images associated with a single source component. The quantity $A$ therefore represents the characteristic astrometric shift induced by the subhalo population and is measured in mas. We performed 1000 realizations for each subhalo mass interval. The resulting distributions of $A$ are presented in Figure~\ref{fig:mass_truncation_criterion}. A clear systematic trend is observed: the median value (red solid lines) and its dispersion of the positional offset decrease as the subhalo populations shift toward lower masses. At the same time, the interquartile range (gray band) becomes progressively narrower, indicating that low-mass subhalos produce weaker and more uniform perturbations. For the largest mass interval $10^{9}$--$10^{10}\,M_\odot$, the median positional offset can reach $A \approx 40$ mas as expected from \cite{Metcalf:2001ap,Chen:2006ym}. For the highest-abundance population, corresponding to the mass interval $10^{5}$--$10^{6}\,M_\odot$, the median positional offset is only $A \approx 0.7789$ mas, while the mean value is $0.7853$ mas. These perturbations lie below the typical $\sim1$ mas astrometric uncertainty of VLBI observations. Consequently, subhalos with masses below approximately $10^{6}\,M_\odot$ are unlikely to generate detectable astrometric shifts in HS~0810+2554. We therefore adopt $M_{\rm min} = 10^{6}\,M_\odot$ as the lower mass threshold in all subsequent subhalo realizations. This choice preserves the dominant astrometric contribution from DM substructure while avoiding the inclusion of a large number of computationally expensive perturbers whose effects remain observationally negligible~\cite{Dalal:2001fq,2014MNRAS.442.2017V}.

Having established the statistical properties of the Aquarius subhalo population and the adopted lower mass threshold, we next examine the lensing effect of an individual DM subhalo through a simplified validation experiment. The purpose of this test is not to construct the final lens model, but rather to demonstrate explicitly that the DM subhalo is capable of producing measurable astrometric perturbations in the HS~0810+2554 system and even resolve the position anomalies. To this end, we introduce a single DM subhalo into the smooth lens and refer to it as the single-subhalo model. The subhalo is represented by a NFW density profile and incorporated into the smooth lens model (EPL profile plus external shear) described previously. Instead of tying the subhalo’s density profile and position to the host halo, we treat the characteristic density $\rho_{s}$, scale radius $r_s$, and projected coordinates $(x_{\rm sub},y_{\rm sub})$ as four free parameters. Therefore, in addition to the free parameters for the smooth model summarized in Table~\ref{tab:mcmc_parameters}, the inclusion of a single NFW subhalo introduces four additional free parameters. These parameters are optimized simultaneously with those of the smooth lens model using the same likelihood function defined by Eq.~\eqref{eq:chi2_pos}. 
The best-fit single-subhalo model yields a noticeable improvement relative to the smooth lens model with a modified $\chi^2 = 30.58$, reducing the positional discrepancies between the predicted and observed image coordinates. The optimized parameters (the fourth column) for the macro-lens are given in Table~\ref{tab:mcmc_parameters}. The four additional parameters are $\rho_{s}=1.59\times10^{16}\,M_\odot\,\mathrm{Mpc}^{-3}$, $r_s=0.433''$ and $(x_{\rm sub},y_{\rm sub})=(0.185'',-0.311'')$. From Table~\ref{tab:mcmc_parameters}, we can see that the best-fit parameters of the smooth and single-subhalo models are very similar, indicating no violation of the macro-lens configuration. The corresponding best-fit subhalo has a mass of $1.261\times10^{10}\,M_\odot$ and the mass of the macro-lens is $1.681\times10^{12}\,M_\odot$. The lensing reconstructions are presented in Figure~\ref{fig:microscopic_analysis} below. This result demonstrates that localized DM perturbations can generate astrometric shifts comparable to those observed in HS~0810+2554. Although a single subhalo is capable of improving the fit, the remaining residuals indicate that the observed anomalies are unlikely to be explained by an isolated perturber alone. This motivates a more realistic subhalo population, where a large number of subhalos contribute simultaneously to the lensing potential as mentioned above. 

\section{RESULTS}

The final stage of the analysis focuses on incorporating Aquarius-based subhalo populations into the macro-lens and systematically exploring the parameter space to identify the optimal lens configuration for the subhalo model. Based on the smooth model, we only keep the Einstein radius and therefore the host halo mass as the free parameter, while all other macro-lens parameters are kept fixed at the best-fit values of the smooth model reported in Table~\ref{tab:mcmc_parameters}. As mentioned above, for each sampled host halo mass, the relevant subhalo mass function is adjusted according to Eq. (\ref{eq:mass_function}). The host-halo mass is sampled over [$1.30\times10^{12}\,M_{\odot}$, $1.70\times10^{12}\,M_{\odot}$], with a spacing of $1\times10^{10}\,M_{\odot}$. Moreover, for each sample of host halo masses, we perform 1000 realizations of the subhalo populations in the way elaborated in \ref{sec:IIec}. 

\begin{figure}[t]
\centering
\includegraphics[width=0.5\textwidth]{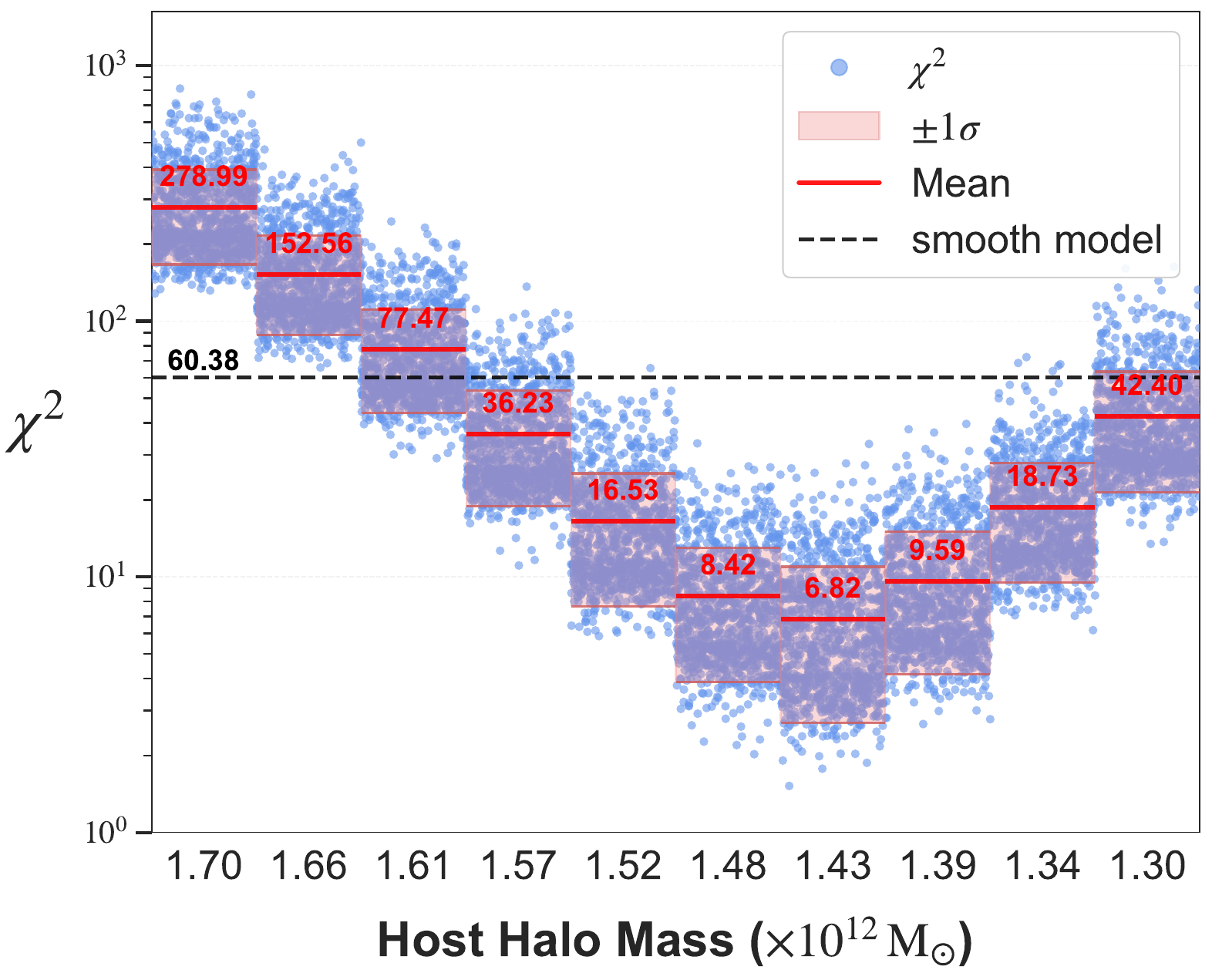}
\caption{Distributions of the $\chi^2$ for the subhalo model for the given host halo masses. The blue scatter points show the $\chi^2$ from 1000 realizations of Aquarius-based subhalo populations across a range of host halo masses. The red curve represents the mean $\chi^2$ and the pink band shows the $\pm1\sigma$ range for each host halo mass. As a reference, the $\chi^2$ for our smooth model (dashed black line) is also shown. The minimum appears at $M_{\text{host}} \approx 1.43 \times 10^{12} M_{\odot}$. This optimal mass configuration minimizes the discrepancy between the substructure-perturbed model and the observed VLBI image positions.}
\label{fig:aquarius_optimization}
\end{figure}

In Figure \ref{fig:aquarius_optimization}, we illustrate the optimization results by showing the relationship between the representative host halo masses and the relevant $\chi^2$ for every 1000 realizations (blue scatter points). We also show the mean $\chi^2$ (red horizontal lines) and $\pm1\sigma$ range (pink bands) for each mass. As a reference, the $\chi^2$ for our smooth model (dashed black line) is also shown and we can see that a large fraction of the realizations lie below it especially around the host halo mass range of [$1.34\times10^{12}\,M_{\odot}$, $1.57\times10^{12}\,M_{\odot}$]. The best-fit solution is obtained for a host halo mass of $1.43\times10^{12}\,M_{\odot}$, with the relevant Einstein radius given by $\theta_E = 0.4257''$ of the EPL plus external shear model. At this mass, the average \(\chi^2\) value across all realizations decreases to 6.82, while the best individual realization reaches \(\chi^2 = 1.61\), showing excellent agreement with the observational data ~\cite{2019MNRAS.485.3009H}. 
Through relaxing the host halo parameters given the best subhalo configuration, the match between the model and the data can be further improved to a minimal $\chi^2$ of 1.18.
We note that, a finer grid of host halo masses was adopted around $M_{\rm host}=1.43\times10^{12}M_\odot$ to better resolve the $\chi^2$ variation in this parameter region. 
The refined scan confirms that $M_{\rm host}=1.43\times10^{12}M_\odot$ corresponds to the minimum of $\chi^2$ over the sampled mass grid and is therefore adopted as the fiducial host halo mass. 


\begin{figure*}[t]
\centering
\includegraphics[width=0.85\textwidth]{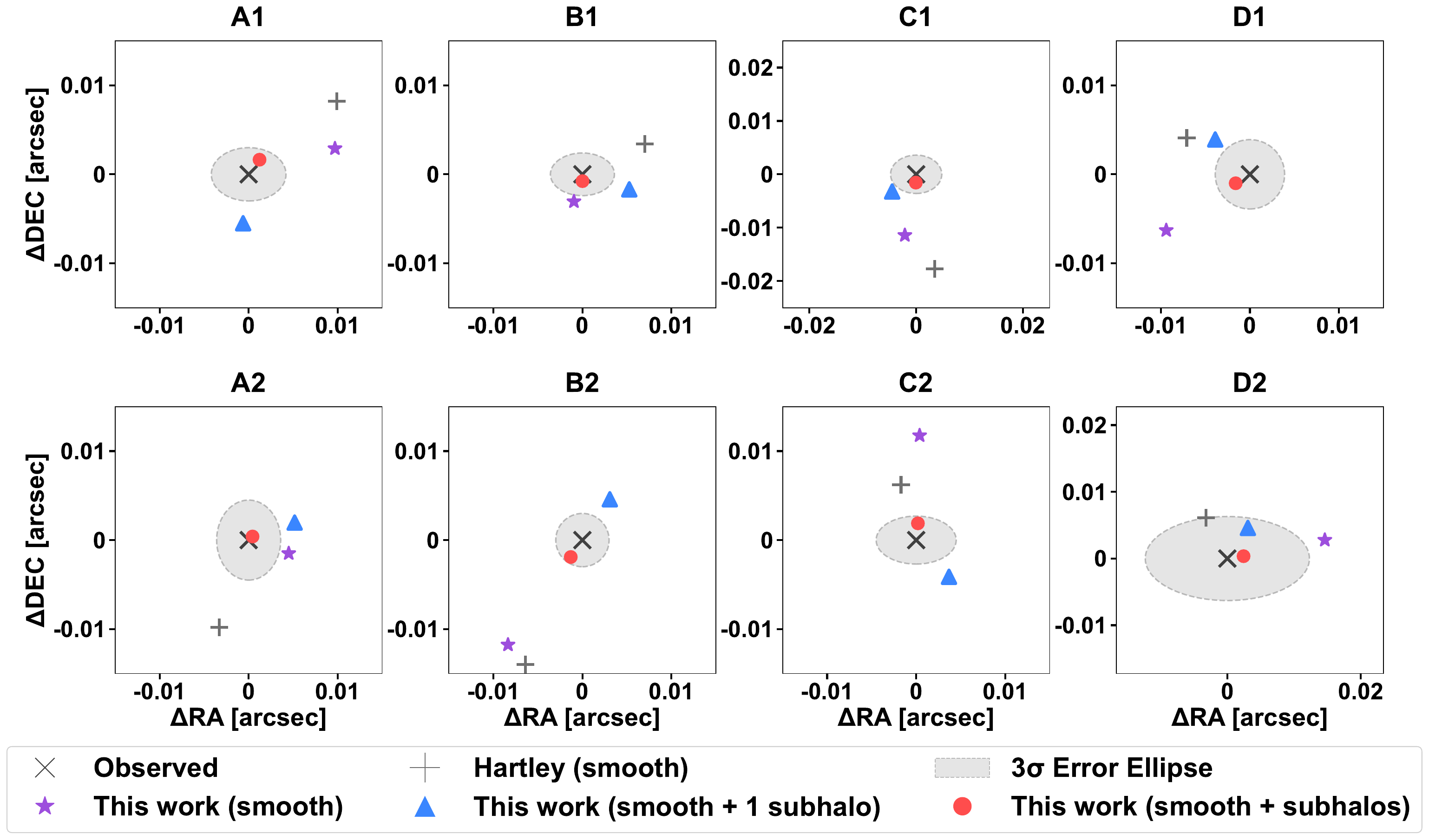} 
\caption{Image-by-image validation of the best-fit model. Each panel provides a mas-scale view of one of the eight lensed images. Observed positions (gray crosses) are compared against predictions from (i) our smooth EPL plus shear model (purple stars), (ii) a single-subhalo model (blue triangles), (iii) our full subhalo model (red circles), and (iv) the smooth model from Hartley et al. (2019) (gray plus signs). Gray ellipses indicate the $3\sigma$ observational uncertainties. The tight clustering of our model predictions within the uncertainty ellipses across all eight images validates the best-fit configuration ($M_{\text{host}} = 1.43 \times 10^{12} M_{\odot}$), achieving an average $\chi^2 = 1.61$.}
\label{fig:microscopic_analysis}
\end{figure*}

The optimized configuration is further validated through an image-by-image comparison of the predicted and observed image positions. Figure \ref{fig:microscopic_analysis} compares the predicted and observed positions of the eight radio images. Each panel provides a mas-scale view of the local image structure. Observed positions are marked by crosses, the predicted image positions of the smooth lens model by stars, the Hartley et al.~model by plus signs, the single-subhalo model by triangles, and the best subhalo realization by circles. The gray ellipses indicate the $3\sigma$ observational uncertainties. The best realization places all eight predicted image positions within, or very close to, the corresponding uncertainty regions, whereas the smooth model retains coherent mas-scale positional residuals. Using the same positional statistic for both models, the minimal $\chi^2$ decreases from 60.38 for the smooth model to 1.61 for the subhalo realizations. This improvement establishes the existence of a CDM-subhalo realization capable of reproducing the octuple-image astrometry in HS 0810+2554. We note that, according to the Aquarius simulations, the cumulative subhalo mass fraction ranges from $6.02\%$ to $13.39\%$ among the six simulated host halos~\cite{2008MNRAS.391.1685S}. The Aq-D-2 halo, whose virial mass $1.774\times10^{12}\,M_\odot$ is most comparable to that of HS~0810+2554, exhibits a cumulative subhalo mass fraction of $13.20\%$. In all of our realizations, the inferred subhalo mass fraction remains below approximately $11\%$, satisfying the subhalo abundance constraints predicted by the Aquarius simulations and remaining consistent with the other predictions of CDM substructure population for galaxy-scale halos~\cite{Jiang:2016yts}.

\section{Discussion}

The failure of a smooth elliptical model to reproduce the VLBI astrometry in HS 0810+2554 establishes the need for additional structure in the lensing potential, but does not uniquely identify its physical origin \cite{2019MNRAS.485.3009H}. Hartley et al.~first interpreted the radio residuals as evidence for a non-smooth mass distribution \cite{2019MNRAS.485.3009H}. Amruth et al.~subsequently showed that FDM-induced density fluctuations can generate positional perturbations of the required scale, while explicitly omitting subhalos from their comparison \cite{2023NatAs...7..736A}. More recently, Miller and Williams demonstrated that higher-order angular multipoles can also substantially reduce the residuals \cite{2025MNRAS.543.3952M}. Previous studies have also demonstrated that strong gravitational lensing is highly sensitive to DM substructure and can provide stringent tests of small-scale structure~\cite{Mao:1997ek,Metcalf:2001ap,Dalal:2001fq}. Our work and results supply the complementary test that was absent from these analyses: a cosmologically motivated population of bound CDM subhalos can itself produce an acceptable reconstruction of the eight radio images without invoking non-standard DM physics.
With its precise astrometry and complex image configuration, HS 0810+2554 therefore remains an important laboratory for testing the predictions of the CDM paradigm on galactic scales~\cite{2020Natur.585...39W}. The mass threshold of $10^{6}\,M_\odot$ identified in our interval-by-interval analysis should be interpreted as an astrometric sensitivity threshold, not as a cutoff in the CDM subhalo mass function. Subhalos below this scale remain part of the assumed population, but their induced positional shifts are smaller than the present VLBI uncertainties. More massive perturbers, particularly those projected near the highly magnified images, dominate the measurable astrometric response. 

Although the Aquarius-based CDM subhalo population can provide a good explanation for the position anomalies observed in HS~0810+2554, several physical effects that may also contribute to lensing perturbations have not been explicitly included in the present analysis. First, our lens model neglects the possible influence of stellar disc within the lens galaxy. Previous studies have shown that edge-on disc can generate lensing perturbations that closely resemble those expected from DM subhalos and may produce significant anomalies in lensing observables even in the absence of a large subhalo population~\cite{Hsueh:2016aih,Hsueh:2017nlk,Hsueh:2017zfs}. Such perturbations may therefore bias the inferred abundance of DM substructure if disc components are not properly modeled. Although there is currently no clear observational evidence for a dominant edge-on disc in HS~0810+2554 \cite{Hsueh:2016aih, 2019MNRAS.485.3009H}, future studies combining high-resolution imaging and stellar-light reconstruction would provide a more robust assessment of possible disc-induced perturbations. Second, the present work adopts a relatively simple EPL plus external shear description for the smooth lens potential. However, real lens galaxies may exhibit departures from idealized elliptical mass distributions, including higher-order angular structures and other forms of mass-model complexity~\cite{2006MNRAS.370.1339H,2018MNRAS.475.1987G,2022MNRAS.516.1808P,2024A&A...688A.110S} as explored by Miller and Williams \cite{2025MNRAS.543.3952M}. Such structures can modify image positions and magnifications and may partially account for residual discrepancies that are otherwise attributed to DM subhalos. Future studies employing more flexible lens models could provide a more complete description of the mass distribution in HS~0810+2554. Third, the present work considers only subhalos associated with the primary lens halo and neglects structures \cite{Chen:2003uu} and DM halos \cite{Metcalf:2004eh} located along the line-of-sight. Gravitational lensing is sensitive to the cumulative mass distribution along the entire light path, and previous studies have demonstrated that line-of-sight halos and intergalactic structure can generate perturbations in image positions that are comparable to those produced by subhalos within the lens galaxy itself~\cite{Xu:2011ru,Xu:2014dda,Hsueh:2019ynk}. Future analyses incorporating cosmologically motivated line-of-sight halo catalogs would provide a more complete description of the lensing model, help distinguish the relative contributions of internal subhalos and intervening halos and potentially achieve better fit to data. 

In addition, although the Aquarius-based subhalo population substantially improves the astrometric reconstruction, our best realization does not necessarily imply that the currently adopted image configuration is unique. Two observational ambiguities may limit the minimum achievable $\chi^2$. First, no time delays have been measured for HS~0810+2554, so the adopted arrival sequence is not directly established observationally \cite{2019MNRAS.485.3009H}. Miller and Williams showed that allowing the reverse ordering can significantly improve the fit, obtaining their overall best solution for a reversed arrival sequence \cite{2025MNRAS.543.3952M}. Second, the association of the eight radio images with the two source components is not unique. Instead, Hartley et al. \cite{2019MNRAS.485.3009H} explored the possible pairings and identified a strongly preferred configuration within their smooth-lens analysis, which we adopt here, whereas the more extensive exploration of Miller and Williams demonstrates that the preferred configuration can depend on the assumed lens complexity and arrival ordering \cite{2025MNRAS.543.3952M}. Exploring these alternatives within our subhalo framework may therefore yield still better reconstructions. A further limitation arises from the astrometric uncertainties themselves. The original VLBI observations resolve internal structure in several radio components, most clearly A2, B2, and C2. Representing such components by single point positions may underestimate the effective source-localization uncertainty; indeed, relaxing the radio astrometric uncertainties to the larger uncertainties of the HST observations substantially improves the fits \cite{2025MNRAS.543.3952M}.

\section{Conclusion}

We have tested whether a cosmologically motivated CDM subhalo population can account for the mas-scale position anomalies of the eight radio images in HS~0810+2554. We find that subhalos below approximately $10^{6}\,M_\odot$ produce shifts below the current astrometric sensitivity, whereas an Aquarius-based subhalo realization above this scale can reduce the positional $\chi^2$ from 60.38 to 1.61. The required subhalo mass fraction is compatible with the Aquarius range. Residual anomalies nevertheless indicate that additional refinements are still required. These include incorporating anisotropic subhalo distributions and accounting for more complex baryonic mass structures explored in related studies. The application of high-resolution simulations together with machine learning techniques~\cite{Hezaveh:2017sht} could also improve the characterization of subhalo populations in complex lens systems like HS 0810+2554. Overall, this work shows that CDM with a cosmologically motivated subhalo population can explain the observed position anomalies in HS 0810+2554 without invoking non-standard DM physics. The modeling framework developed here can be directly extended to other quadruply imaged strong lens systems with comparable astrometric precision. Future studies combining improved lens modeling with additional lensing observables will provide more stringent constraints on the origin of these anomalies and on the nature of DM.

\acknowledgments 
This work is supported by the National Natural Science Foundation of China (No. 12220101003) and the Project for Young Scientists in Basic Research of Chinese Academy of Sciences (No. YSBR-061).

\bibliography{reference}

\end{document}